\documentclass[fleqn,twoside]{article}
\usepackage{espcrc2}

\usepackage{epsfig}
\usepackage[dvips]{color}
\usepackage{latexsym}

\title{Static hybrid quarkonium potential with improved staggered quarks
}
\author{ MILC Collaboration: C.~Bernard
\address{Department of Physics, Washington University, St.~Louis, MO 63130, USA},
T.~Burch
\address{Department of Physics, University of Arizona, Tucson, AZ 85721, USA}, 
C.E.~DeTar
\address{Physics Department, University of Utah, Salt Lake City, UT
  84112, USA},
Ziwen Fu$\,\null^{\rm c}$\thanks{Presented by Ziwen Fu},
Steven~Gottlieb
\address{Department of Physics, Indiana University, Bloomington, IN 47405, USA and Fermilab, Batavia, IL 60510, USA},
E.~Gregory$\,\null^{\rm b}$,
U.M.~Heller
\address{CSIT, Florida State University, Tallahassee, FL 32306-4120, USA},
J.~Osborn$\,\null^{\rm c}$,
R.L.~Sugar
\address{Department of Physics, University of California, Santa Barbara, CA 93106, USA},
and D.~Toussaint$\,\null^{\rm b}$
} 

\begin{document}

\begin{abstract}

We are studying the effects of light dynamical quarks on the
excitation energies of a flux tube between a static quark and
antiquark.  We report preliminary results of an analysis of the ground
state potential and the $\Sigma^{\prime+}_g$ and $\Pi_u$ potentials.
We have measured these potentials on closely matched ensembles of
gauge configurations, generated in the quenched approximation and with
2+1 flavors of Asqtad improved staggered quarks.

\end{abstract}

\maketitle

\section{INTRODUCTION}

Simulations with dynamical quarks have found that light quarks modify
the heavy quark-antiquark potential in a number of ways
\cite{Bernard_034503,Wilson_dynamical,UKQCD}.  At large distances they
decrease the string tension in units of the Sommer $r_0$ and $r_1$
parameters (defined by $r^2 F(r) = 1.65$ and 1.00, respectively) and
lead eventually to string-breaking.  At shorter distances they modify
the running of the coupling constant, deepening the Coulomb well and
increasing the ratio $r_0/r_1$. In this work, we extend these studies
to some of the potentials with excited flux tubes.  Of particular
interest to quarkonium spectroscopy are the $\Pi_u$ excitations
leading to exotic $Q\bar Qg$ hybrids \cite{JKM}.

We report results of a study in which our sources and sinks are
optimized to create and annihilate a flux-tube state.  In the presence
of dynamical quarks, string breaking is expected.  It is known that in
the conventional $\Sigma_g^{+}$ channel, transitions to the open
two-meson channel are exceedingly weak, qualitatively consistent with
the small widths of quarkonium states above the heavy-light meson
thresholds \cite{StringBreak,UKQCD}.  Since at present we do not include the
open two-meson channel we do not expect to observe string breaking
here.

\section{MEASUREMENTS} 
We have measured the heavy quark potential on an ensemble of
$28^3\times 96$ ($a \approx 0.09$ fm) gauge configurations generated
in the presence of $2+1$ flavors of Asqtad dynamical quarks of varying
masses and a one-loop Symanzik gauge action\cite{Asqtad}.  The strange
quark mass is set approximately to its physical value.  Here we
compare results from our 358-configuration quenched ensemble with our
495-configuration dynamical quark ensemble for which $(m_\pi r_0)^2
\approx 1.3$.

The configurations are first smoothed using a single hypercubic (HYP)
blocking pass \cite{Hasenfratz_Knechtli}, a technique that improves
significantly the signal-to-noise ratio \cite{Hasenfratz_Hoffmann}.
The blocking procedure involves replacing all gauge links (timelike as
well as spacelike) with an SU(3)-projected average over paths confined
to adjacent hypercubes.  Thus distortions in the result are local and
expected to be confined to distances smaller than about $2a$
\cite{Hasenfratz_Hoffmann}.  After HYP blocking the spacelike links
are further smoothed via five cycles of APE smearing with SU(3)
projection.

\begin{figure}[t]
 \hspace*{-12mm}
 \epsfig{bbllx=100,bblly=230,bburx=730,bbury=740,clip=,
         file=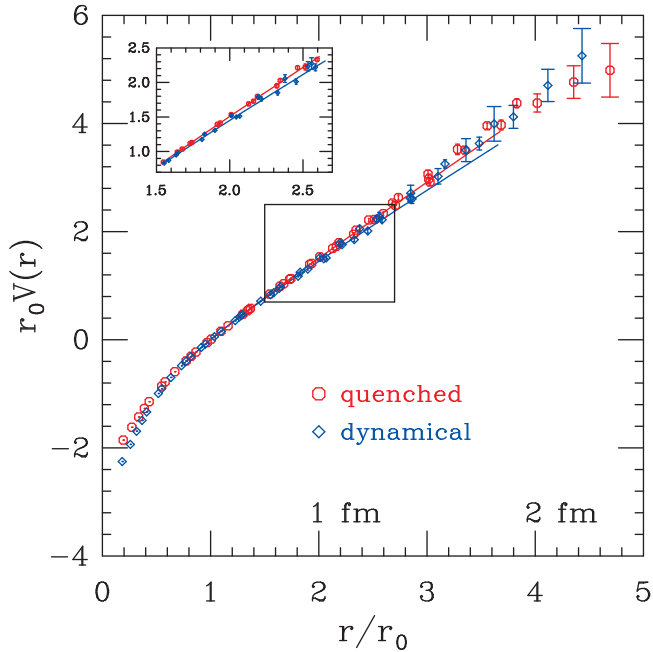,width=140mm} \\
 \vspace*{-35mm}
\caption{ 
\label{GroundPotential}
The ground state static quark potential for quenched (octagons) and
2+1 flavor (diamonds) QCD, in units of $r_0$.  The solid lines are
fits to the Coulomb plus constant plus linear form, fixing $V_{\rm
fit}(r_0) = 0$. The lattice spacing is matched using $r_0$.  The inset
expands the area shown by the box.  }
 \vspace*{-5mm}
\end{figure}

On the thus smoothed lattices we measure the expectation value of the
standard $R\times T$ Wilson loop on axis and along three different
off-axis directions.  These measurements yield the conventional ground
state $\Sigma_g^+$ and excited state $\Sigma_g^{+\prime}$ potentials.
For the $\Pi_u$ excited state, we measured the expectation value of a
bent loop formed by replacing the source and sink flux tubes of length
$R$ by a superposition of large ``staples'' of sides $(2a, R, 2a)$.
For example one such loop replaces each on-axis spacelike flux path
($R\hat x$) by paths of the form ($2a \hat y$, $R\hat x$, $-2a \hat
y$) minus its reflection in the $xz$ plane \cite{GMR}.

For the standard Wilson loop we extracted the usual $\Sigma_g^+$
potential $V_{\Sigma g+}$ and its excited state $V_{\Sigma g+}^\prime$
by doing a blocked, correlated, double-exponential fit to
the Wilson loop data:
\begin{eqnarray}
  W(R,T) &=& C_{\Sigma g+}(R)e^{-V_{\Sigma g+}(R)T} \\
    &+& C_{\Sigma g+}^\prime(R)e^{-V_{\Sigma g+}^\prime(R)T}.
\end{eqnarray}
For the $\Pi_u$ potential we did only a single-exponential fit.

In all cases we use the same fit ranges for both quenched and dynamical
lattices to reduce possible systematic errors.

\begin{figure}[t]
 \hspace*{-12mm}
\epsfig{bbllx=100,bblly=230,bburx=730,bbury=740,clip=,
         file=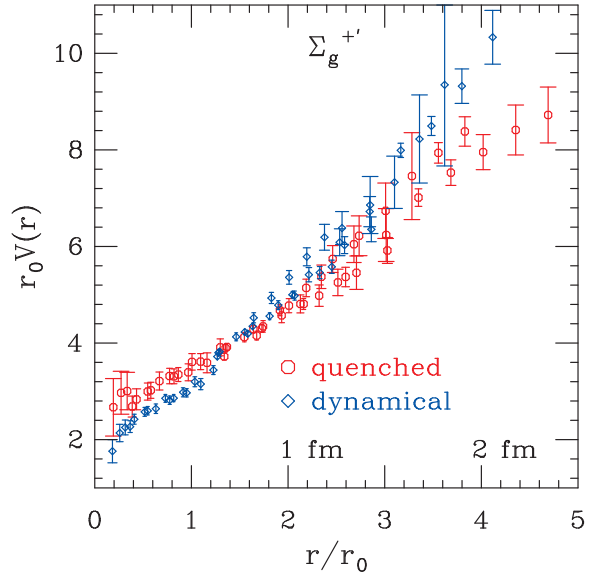,width=125mm} \\
\vspace*{-35mm}
\caption{ 
\label{Excited_Potential}
The  excited $\Sigma_g^\prime$ potential for quenched (octagons) 
and 2+1 flavor (diamonds) QCD, in units of $r_0$.   
The lattice spacing was matched using $r_0$. }
\vspace*{-5mm}
\end{figure}

\section{RESULTS}

In Fig.~\ref{GroundPotential} we compare the ground state potential on
the quenched ensemble and the 2+1 flavor ensemble.  Both the distance
scale and the potential are plotted in units of $r_0$, and a constant
has been subtracted from the potential so that it is zero at $r_0$.
Since $r_0$ was determined from this potential, the fits are tangent
at this point.  Away from $r_0$, the potentials have different shapes,
namely, the Coulomb attraction is slightly stronger in the light quark
ensemble and the string tension is slightly weaker in units of $r_0$,
confirming earlier findings \cite{Bernard_034503}.  A softening of the
Coulomb well is also evident.  This is an expected consequence of HYP
smoothing.

Similarly in Fig.~\ref{Excited_Potential} we show light quark effects
in the $\Sigma_g^\prime$ excitation potential.  The potentials are
plotted relative to the zero determined in the fit to the ground state
potentials.  From Fig.~\ref{Excited_Potential} we see that the
excited state potential $\Sigma_g^\prime$ is slightly steeper than
that of quenched QCD.

\begin{figure}[t]
 \hspace*{-12mm}
\epsfig{bbllx=100,bblly=230,bburx=730,bbury=740,clip=,
         file=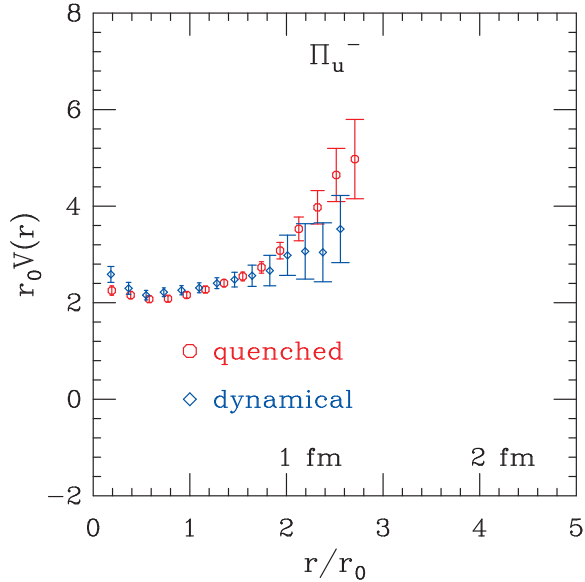,width=125mm} \\
\vspace*{-35mm}
\caption{
\label{Pi_u_Potential} 
The $\Pi_u^-$ potential for quenched (octagons) and 2+1 flavor
(diamonds) QCD, in units of $r_0$. The potentials are plotted relative
to the zero determined in the fit to the ground state potentials.}
\vspace*{-8mm}
\end{figure}

In Fig.~\ref{Pi_u_Potential} we plot the $\Pi_u^-$ potential.  This
hybrid potential is weakly repulsive at short range, as would be
expected from the Coulomb interaction in a color octet quark-antiquark
system.  This effect is softened by HYP smoothing.  At the level of
our statistical errors there are no apparent differences at long
range, but better statistics would certainly be of interest.
\section{CONCLUSIONS}

Our measurements at $a = 0.09$ fm confirm the shape changes in the
ground-state potential, seen previously at $a = 0.13$ fm.  In units of
$r_0$ we find, further, that adding $2+1$ flavors of dynamical quarks
makes the $\Sigma_g^{+\prime}$ excited state potential slightly
steeper and the $\Pi_u^-$ slightly more repulsive at short range.  We
find no clear evidence for a flattening of the potentials that would
signal string breaking.

Computations were performed at LANL, NERSC, NCSA, ORNL, PSC, SDSC,
FNAL, and the CHPC (Utah).  This work is supported by the U.S. NSF and
DOE.

%


\begin{thebibliography}{20}
\bibitem{Bernard_034503} 
C.~Bernard {\it et al.},
Phys.\ Rev.\ D {\bf 62} (2000) 034503
and D.~Toussaint, this conference (2002).
\bibitem{Wilson_dynamical}  
G.~S.~Bali {\it et al.}  [T$\chi$L Collaboration],
Phys.\ Rev.\ D {\bf 62} (2000) 054503.
C.~Allton  [UKQCD Collaboration],
Nucl.\ Phys.\ Proc.\ Suppl.\  {\bf 109} (2002) 3.
\bibitem{UKQCD}
B.~Bolder {\it et al.},
Phys.\ Rev.\ D {\bf 63} (2001) 074504.
\bibitem{JKM}
K.~J.~Juge, J.~Kuti and C.~J.~Morningstar,
Phys.\ Rev.\ Lett.\  {\bf 82} (1999) 4400.
and
Nucl.\ Phys.\ Proc.\ Suppl.\  {\bf 83} (2000) 304.
\bibitem{StringBreak}
C.~Bernard {\it et al.},
Phys.\ Rev.\ D {\bf 64} (2001) 074509.
I.~T.~Drummond and R.~R.~Horgan,
Phys.\ Lett.\ B {\bf 447} (1999) 298.
\bibitem{Asqtad} 
%
K.~Orginos and D.~Toussaint,
Phys.\ Rev.~\ D {\bf 59} (1999) 014501;
%
K.~Orginos, D.~Toussaint and R.~L.~Sugar,
Phys.\ Rev.~\ D {\bf 60} (1999) 054503;
%
G.~P.~Lepage,
Phys.\ Rev.\ D {\bf 59} (1999) 074502.
\bibitem{Hasenfratz_Knechtli} 
%
A.~Hasenfratz and F.~Knechtli,
Phys.\ Rev.~\ D {\bf 64} (2001) 034504.
\bibitem{Hasenfratz_Hoffmann} 
A.~Hasenfratz, R.~Hoffmann and F.~Knechtli,
Nucl.\ Phys.\ Proc.\ Suppl.\  {\bf 106} (2002) 418.
\bibitem{GMR}
L.~A.~Griffiths, C.~Michael and P.~E.~Rakow,
Phys.\ Lett.\ B {\bf 129} (1983) 351.
\end{thebibliography}
\end{document}